\documentclass[useAMS,usenatbib]{mnras}
\usepackage{natbib}
\usepackage{aas_macros}
\usepackage{graphicx}
\usepackage{savesym}
\usepackage{listings}
\usepackage[intlimits]{amsmath}
\savesymbol{iint}
\usepackage{txfonts}
\usepackage{bm}
\restoresymbol{TXF}{iint}
\usepackage{amssymb}
\usepackage{hyperref}
\usepackage{amsmath}
\usepackage{tabularx} 
\usepackage{color}
\usepackage{soul}
\usepackage{ulem}
\newcommand{\cE}{{\cal E}}

\newcommand{\vv}[1]{\bmath{#1} }

\newcommand{\spr}[2]{\bmath{#1} \!\cdot\! \bmath{#2}}
\newcommand{\vpr}[2]{\bmath{#1} \!\times\! \bmath{#2}}
\newcommand{\vdiv}{\bmath\nabla\!\cdot\!}
\newcommand{\vgrad}[1]{\nabla{#1}}
\newcommand{\vcurl}[1]{\vpr{\nabla}{#1}}

\newcommand{\oder}[2]{\frac{d #1}{d #2}}

\newcommand{\pder}[2]{\frac{\partial #1}{\partial #2}}

\newcommand{\Pd}[1]{\partial_{#1}}

\newcommand{\ort}[1]{ \bmath{e}_{#1} }

\newcommand{\sub}[1]{_{\mbox{\tiny #1}}}
\newcommand{\beq}{\begin{equation}}
\newcommand{\eeq}{\end{equation}}

\def\A{Alfv\'en }

\def\tom{\tilde{\omega}}
\def\tk{\tilde{k}}

\def\apj{Astrophys. J.}
\def\apjl{Astrophys. J. Lett.}
\def\apjs{Astrophys. J. Supp. Ser. }

\def\aap{Astron. Astrophys. }

\def\jcp{J. Comput. Phys. }
\def\mnras{Mon. Not. Roy. Astron. Soc. }

\def\pre{Phys. Rev. E.}

\title[Parametric instability of Alfv\'en waves]{Parametric instability of Alfv\'en wave packets}
\author[S. S. Komissarov] 
{
Serguei S. Komissarov \thanks{E-mail: s.s.komissarov@leeds.ac.uk}\\
School of Mathematics, The University of Leeds, Leeds, LS2 9JT, UK}

\begin{document}
\date{Received/Accepted}
\maketitle

\begin{abstract} 
Parametric instability of Alfv\'en wave packets with monochromatic carrier wave in low-$\beta$ plasma is studied using one-dimensional magnetohydrodynamic simulations. The results show spatial growth of incoming perturbations as they propagate through the mother wave. For sufficiently short packets, the perturbations emerge downstream of the packet as small-amplitude reverse Alfv\'en waves and forward slow magnetosonic waves. For larger packets the perturbations reach non-linear amplitude while still inside the mother wave.  In this case, a downstream section of the mother wave collapses but the remaining upstream section stays largely intact and enters the phase of very slow evolution. The length scale separating the linear and non-linear regimes, as well as determining the size of the surviving section in the non-linear regime, is set by the Alfv\'en crossing time of the packet, the growth rate of the parametric instability for the unmodulated carrier wave, and the amplitude of incoming perturbations.  The results are discussed in connection with the physics of solar wind.    
\end{abstract}   
                                                                                       

\section{Introduction}
\label{introduction}

Both the solar corona and its wind are subject to heating via some nonthermal mechanism. The solar corona is extremely hot compared to Sun's photosphere, and the temperature of solar wind decreases much slower than predicted in its adiabatic model \citep{GBML94}. Heating by plasma waves emitted by the dynamic Sun's surface has been considered as one of such mechanisms. Compressive MHD modes, like fast and slow magnetosonic modes may rapidly convert their energy into heat via non-linear steepening followed by development of shock waves \citep[e.g.][]{Whitham,CH74} or collisionless Landau damping \citep[e.g.][]{MT64,Barnes66,Kellogg20}. However, the effectiveness of their dissipation is also a drawback, as they cannot reach required large distances from the Sun. In contrast, the phase speed of \A waves does not depend on their amplitude and so they do not steepen into shocks. They are also much less affected by the collisionless dissipation.  So in principle, they can propagate large distances without attenuation, and the interplanetary missions reveal that apart from the fast streams (advective waves), the solar wind perturbations are dominated by \A waves.  However, the effectiveness of \A waves at energy transport may turn into a disadvantage unless a  suitable mechanism is found for converting their energy into heat. It is believed that the parametric instability of large-amplitude \A wave may be such a mechanism. 

The instability was discovered in 1960s by \citet{GO62,SG69}, who studied the stability a small-amplitude (linear) circularly-polarised monochromatic \A wave (mother wave) to even smaller perturbations in the framework of one-dimensional magnetohydrodynamics (1D MHD).  To simplify the analysis, they assumed high plasma magnetisation, which allowed them to identify and then retain only the dominant coupling term in the perturbation equations.  Under these conditions the perturbations are sound and \A waves (daughter waves). The instability develops under resonance conditions involving wave numbers and frequencies.  \citet{SG69} also pointed the similarity between this problem and the resonance interaction between three coupled harmonic oscillators.

Later, \citet{Gold78} and \citet{Derby78} independently analysed this stability problem without using the simplifications of \citet{SG69}. The daughter waves are no-longer sound and \A waves but only resemble them.  The original results by \citet{SG69} are recovered in the limit $\eta^2\ll b^2 \ll 1$ \citep{Derby78}, where $b=a/c_0$ is the ratio the sound speed $a$ and the \A speed along the wave vector, $c_0=B_0/4\pi \sqrt{\rho_0}$, there $B_0$ is the magnitude of magnetic field component along the wave vector, and $\rho_0$ is the unperturbed plasma density, and $\eta=||\vv{B}_{\perp,0}||/B_0>0$ is the ratio of the magnitudes of the transverse and longitudinal components of the magnetic field in the mother wave.  \citet{JH93} carried out detailed analysis of the solutions and derived analytical approximations for the unstable modes.    \citet{RS04} used an alternative approach to this problem, by reducing the original system of perturbation equations, which contains periodic coefficients, to a system with constant coefficients.       

In the nonlinear phase, which has been studied with computer simulations, the non-linear steepening of longitudinal perturbations  leads to formation of shock waves, resulting in efficient heating of plasma \citep[e.g.][]{DZ01a}.  The parametric instability is also considered as a driver for \A turbulence in the solar wind (see \citet{BC13review} and the references there, \citet{SY18}), which transfers energy to small scales where it dissipates via collisionless mechanisms.  

The magnetic and velocity perturbations of the solar wind are dominated by outgoing \A waves, but the observations also suggest a significant contribution by the ingoing waves \citep[e.g.][]{GSB95,BBP96}, and the generation of  ingoing waves via the parametric instability of the outgoing ones is an attractive mechanism.   The numerical simulations have been naturally extended from 1D to 2D and 3D to explore effects associated with the increased dimensionality \citep[e.g.][]{DZ01,DZ15,SY18,PMS19}     

The investigation of the parametric instability has been extended to arc-polarised \citep{DZ01a, MT24}, weakly non-monochromatic waves \citep{MV96}, and to the frameworks of relativistic MHD \citep{II24}, two-fluid  \citep[e.g.][]{VG91}, hybrid \citep[e.g.]{}, and fully kinetic models \citep[e.g.]{GIT23}. Recently, \citet{MT24} extended the investigation to \A wave packets. 

Arc-polarised wave packets have been identified in the data obtained with interplanetary probes \citep{LS80,Tsu96,Riley96}. According to the statistical analyses by \citet{Riley96}, essentially all such waves move away from the Sun in the rest frame of the solar wind plasma, which suggests that they are originated at or near the Sun. They account for 5 to 10\%  of the Ulysses data.  The magnetic field rotation is limited to $180^\circ$ with no preferred helicity. So, the question whether such packets decay via the parametric instability, and if they do then how rapidly, is very relevant to the physics of solar wind.  

According to the semi-analytical study by \citet{MT24}, who solved numerically the boundary value problem for the eigenmodes of the instability, such packets are still subjects to the parametric instability, though with a weaker growth rate compared to arc-polarised monochromatic waves. Moreover, the growth rate scales like $\simeq l/L $ when $l/L\to 0$, where  $l$ the linear size of the packet and  $L$ is the size of the computational domain.   This property seems rather bizarre, particularly in the case of open boundary conditions (BCs), because the wave interaction is fully local and hence should be limited to the region occupied by the packet. Therefore the growth rate should not depend on $L$ when $L>l$.          

In this paper we study the parametric instability of \A wave packets using one-dimensional ideal MHD simulations. Section \ref{sec:theory} provides the relevant theoretical background to the problem.   Section \ref{sec:study} describes the method and the plasma parameters common to all the simulations described in this paper.  In section \ref{sec:mono-waves}, we present the results for monochromatic circularly- and arc-polarised \A waves, which provides a reference point for the main study of  arc-polarised wave packets described in section \ref{sec:apwp}.   Section \ref{sec:discussion} contains the general discussion of these results and their implications.

\section{Theoretical background} 
\label{sec:theory}

In their analysis, \citet{SG69} assume the transverse magnetic field of the mother \A wave in the form $\vv{B}_{0,\perp} = \vv{A}\exp(i(k_0z -\omega_0t)) + \vv{A}^*\exp(-i(k_0x -\omega_0t))$, where $\vv{A}$ is a complex amplitude vector. Although in the text they describe the mother wave as circularly-polarised, this a most general representation of a monochromatic wave which allows all types of polarisation, from linear to circular, depending on $\vv{A}$, which is not specified in their work. The same description is used for daughter waves.  Then they  show that the amplitudes of the daughter waves can have secular evolution (on the time scale longer than the periods of their oscillations) under the resonance conditions 
\beq
\omega_+=\omega_0 + \omega\,,\quad k_+=k_0+k\,.
\label{eq:resonance+}
\eeq                 
Here $\{\omega_0,k_0\}$, $\{\omega_+,k_+\}$, $\{\omega,k\}$ are the real frequencies and wave numbers of the mother \A wave, the daughter \A wave, and the sound wave respectively\footnote{The notation is modified to align with the rest of the paper.}.  To simplify the discussion, and without loss of generality, they assumed $\omega_0=c_0 k_0$, where $c_0=B_0/\sqrt{\rho_0}>0$ is the \A speed in the direction of the phase vector\footnote{Note the disappearance of the factor $1/\sqrt{4\pi}$ via renormalisation  of the magnetic field here and throughout this paper.}. This describes a forward wave in the rest frame of plasma. Both positive and negative values of $k_0$ and $\omega_0$ are allowed.  With more accurate description, these two options correspond to opposite polarisations and helicities of the mother wave.   

This circular evolution of the daughter waves is an exponential growth when the \A daughter wave travels in the opposite (reverse) direction to the mother wave, $\omega_+/k_+=-c_0$, and the sound wave travels in the same (forward) direction as the mother wave, $\omega=a k$. 

Quick inspection of equations (I-17,I-18) in \citep{SG69}, after accounting for several typos,  shows that three more resonances are also possible. So, altogether there four resonances which can be combined into two pairs,      
\beq
\omega_\pm=\omega \pm \omega_0\,,\quad k_\pm=k\pm k_0\,,
\label{eq:res-p1}
\eeq
and
\beq
\bar{\omega}_\pm=-\omega \pm\omega_0\,,\quad \bar{k}_\pm=-k \pm k_0\,.
\label{eq:res-p2}
\eeq
Since $\omega_\pm/k_\pm=\bar{\omega}_\mp/\bar{k}_\mp$, the pairs have the same phase speeds.  Repeating the calculations of \citet{SG69}  for the new resonances yields basically the same results. For each of the resonances, the instability occurs when the \A daughter is a reverse wave, and the sound wave is a forward wave,  with the same growth rate in all cases.   

\citet{Gold78} and \citet{Derby78} start their analysis by introducing the mother \A wave as a proper circularly-polarised transverse wave. To this aim, \citet{Gold78} uses complex Jones vectors, whereas \citet{Derby78} introduces complex variables for the transverse components of the magnetic field $B_{\pm} = B_{y} \pm i B_{z}$ and $U_{\pm} = U_{y}\pm i U_z$. For a circularly-polarised monochromatic transverse wave, 
\beq
B_+= \hat{B} e^{i\phi}\,,\quad B_-=\hat{B}^* e^{-i\phi^*} \,,
\eeq
where the phase $\phi=kx-\omega t$, and the complex amplitude $\hat{B}=B_\perp e^{i\phi_s}$. To incorporate the possible temporal exponential growth, the frequency is assumed to be complex, $\omega=\omega_r+i\gamma$, where $\omega_r\in R$ is the proper frequency and $\gamma\in R$ is the growth rate. $B_\perp\in R$ is the strength of the transverse magnetic field, $k\in R$ is the wave number, and $\phi_s\in R$ is the constant phase shift.  

For the mother \A wave, $\gamma=0$, and without any loss of generality one may assume $\phi_s=0$.  Hence, for its transverse magnetic field,  
 $B_{0,\pm}=B_{0,\perp} \exp(\pm i(k_0x-\omega_0 t))$. For the corresponding transverse velocity  $U_{0,\pm}=U_{0,\perp} \exp(\pm i(k_0x-\omega_0 t))$, where $\omega_0/k_0=c_0$, and $U_{0,\perp}=B_{0,\perp}/\sqrt{\rho_0}$, where $\rho_0$ is the unperturbed uniform plasma density. 

The system of linearised equations for the perturbations $\vv{u}$, $\vv{b}$, and $\rho$ is solved using the method of Fourier transform.  The calculations reveal that, with one exception, its normal modes are not normal in the sense that they are not associated with a single frequency and a single phase speed, but describe a coexistence (symbiosis) of three sub-waves whose frequencies and wave numbers satisfy the resonance conditions \eqref{eq:res-p1}.  One of the sub-waves is longitudinal and controls the variations of gas density, pressure, and the longitudinal component of velocity. The other two sub-waves are transverse, as they control the variations of the transverse components of the magnetic field and velocity. 

If the longitudinal sub-wave is described by the harmonic 
\beq 
\rho = \hat{\rho}e^{i\phi} + \hat{\rho^*}e^{-i\phi^*}\,,
\eeq 
where $\phi=k x -\omega t$, then the perturbation equations dictate 
\beq
\begin{split}
u_x &= \hat{u}_x e^{i\phi} + \hat{u}^*_x e^{-i\phi^*} \,,\\
b\pm &= \hat{b}_{\pm} e^{i\phi_{\pm}} + \hat{b}^*_{\mp} e^{-i\phi^*_{\mp}}\,, \\
u\pm &= \hat{u}_{\pm} e^{i\phi_{\pm}} + \hat{u}^*_{\mp} e^{-i\phi^*_{\mp}}\,, \\
\end{split}
\eeq
where $\phi_{\pm}=k_{\pm} x -\omega_{\pm} t$. $k_\pm$ and $\omega_\pm$ are still given by the resonance conditions \ref{eq:res-p1}, but the frequencies are now allowed to be complex.  It is easy to see that the transverse perturbation is a sum of two circularly-polarised sub-waves,  one with 
\beq
\begin{split}
& b_+=\hat{b}_{+} e^{i\phi_{+}}\,,\quad b_-=\hat{b}_{+}^* e^{-i\phi_{+}^*}\,,\\ 
& u_+=\hat{u}_{+} e^{i\phi_{+}}\,,\quad u_-=\hat{u}_{+}^* e^{-i\phi_{+}^*}\,,\\
\end{split}
\eeq
and the other with  
\beq
\begin{split}
& b_+=\hat{b}_{-}^* e^{-i\phi_{-}^*}\,,\quad b_-=\hat{b}_{-} e^{i\phi_{-}}\,,\\ 
& u_+=\hat{u}_{-}^* e^{-i\phi_{-}^*}\,,\quad b_-=\hat{u}_{-} e^{i\phi_{-}}\,,\\
\end{split}
\label{eq:corrected}
\eeq
({\color{red}In the version published in MNRAS, the exponents in equation \eqref{eq:corrected} have wrong sings.}) 
The complex amplitudes of the sub-waves are related via

\beq
\begin{split}
\tilde{u}_x &= (\tom/\tk)\tilde{\rho} \,,\\
\tilde{b}_\pm &= \eta\frac{ (\tk \pm 1)(\tom^2 \pm (2\tom -\tk) }{\tk (\tom - \tk)(\tom+\tk\pm 2)} \tilde{\rho} \,,\\
\tilde{u}_\pm &= -\eta\frac{  (\tom\tk^2 + 2\tom -\tk) \pm(\tom^2 -2\tom\tk -\tk)   }{\tk (\tom - \tk)(\tom+\tk\pm 2)} \tilde{\rho} \,,\\
\end{split}
\label{eq:amplitudes}
\eeq
where $\tilde{\rho}=\hat{\rho}/\rho_0$, $\tilde{u}=\hat{u}/c_0$,  $\tilde{b}=\hat{b}/B_0$, $\tk=k/k_0$, $\tom=\omega/\omega_0$ \citep{Derby78}.  

The phase speeds are given by the dispersion equation\footnote{Both in \citep{Gold78} and \citep{Derby78}, the common factor $(\tom-\tk)$ is dropped as it is not relevant to the instability. However, it is still important for the complete description of the normal modes of the system.} 
\beq
   \begin{split}
      (\tom-\tk)& \Big[ ((\tom+\tk)^2-4)(\tom-\tk)(\tom^2- b^2 \tk^2) - \\
         \qquad&  -\eta^2\tk^2(\tom^3+\tk\tom^2 -3\tom +\tk)\Big]=0 \,,\\
   \end{split}\
   \label{eq:dispersion}
\eeq  
where $b=a_0/c_0$, where $a_0$ is the sound speed of the unperturbed state. 
One root of \eqref{eq:dispersion} is the solution to $(\tom-\tk)=0$. The corresponding phase speed is $\omega/k=c_0$, the same as the phase speed of the mother wave. Since the factor $(\tom-\tk)$ also appears in the denominator for the last two expressions in \eqref{eq:amplitudes}, this root describes a pure transverse mode. Quick inspection of the perturbation equations with vanishing density perturbation show that pure transverse perturbations are allowed provided they have the same phase function as the mother wave, $b_\pm = \hat{b}_\pm\exp(\pm i(k_0x-\omega_0t))$.  Thus, this solution amounts to variation of the   amplitude and phase shift of the mother wave.   The other roots correspond to proper symbiotic modes.  Strictly speaking, the symbiotic modes are dispersive, as their phase speed $v_{p}=\omega/k$ depends on the wave number.       
      
Equation \eqref{eq:dispersion} is symmetric with respect to the transformation $\tk\to-\tk$, $\tom\to-\tom^*$. Since $\{\tk,\tom\}$ and $\{-\tk,-\tom^*\}$ describe the same longitudinal sub-wave, the same solutions appear twice on the dispersion diagram, and hence one may limit their analysis to $k\ge0$. Figure \ref{fig:disp} illustrates the dispersion diagram for the case with $b^2=0.1$ and $\eta=1$, explored in the simulations described in sections \ref{sec:mono-waves} and  \ref{sec:apwp}.  One can see that it is quite complicated and involves a number of interesting bifurcations. The most important ones describes the transitions between stable and unstable modes. Everywhere, except the region $1.21<\tk<3.43$, all the modes are stable and so one can count exactly six different real solutions of the dispersion equation for $\tom$. In the unstable region, there are two complex conjugate roots of the dispersion equation, and in the plot of $\omega_r$  against $\tk$  one can count only five different dispersion curves.  One may describe the instability as a resonance interaction between between two symbiotic modes when they have the same values of $k$ and $\omega_r$, and hence the same $k_\pm$ and $\omega_{\pm,r}$. 

The original results by \citet{SG69} are recovered in the limit  $\eta^2\ll b^2 \ll 1$ \citet{Derby78}.  In this limit, the solutions to the dispersion equation are approximately the double root to $(\tom-\tk)^2=0$, the two roots to $((\tom+\tk)^2-4)=0$, and the two roots to $(\tom^2- b^2 \tk^2)=0$. For the first four roots, the denominators in the last two equations \eqref{eq:amplitudes} vanish suggesting degeneration of corresponding symbiotic modes into purely transverse waves. Since these roots yield the phase speed $v_{p}=\pm c_0$, these are \A waves \citep{JH93}. For the last two roots, $v_{p}=\pm a$, the denominators in \eqref{eq:amplitudes} do not vanish, and the small value of $\eta$ implies degeneration of the corresponding symbiotic modes into the purely longitudinal sound waves.  The instability occurs at the intersection point of the lines $\tom+\tk-2=0$ and $\tom = b\tk$, which describe the reverse \A wave and forward sound wave respectively.  In the general case with $b<1$, this point stretches into a line segment, like the dashed line in figure \ref{fig:disp}  \citep{JH93,RS04}.

\section{The method} 
\label{sec:study}

In this study, we numerically solve the equations of compressible ideal MHD in the form of conservation laws, which include the continuity equation

\beq
\pder{\rho}{t}+ \nabla\!\cdot\!(\rho \vv{U}) =0\,,
\eeq
the Euler equation
\beq
\pder{\rho \vv{U}}{t}+ {\nabla}\cdot (\rho \vv{U}\otimes \vv{U}-\vv{B}\otimes\vv{B}) + 
\nabla\left(p+\frac{||\vv{B}||^2}{2}\right)=0\,,
\eeq
the energy equation 
\beq
   \begin{split}
    &\pder{}{t}\left(\frac{\rho || \vv{U}||^2}{2}+e_{th} + \frac{||\vv{B}||^2}{2}\right) +\\
     &\qquad+ \nabla\cdot\left[ \left(\frac{\rho || \vv{U}||^2}{2}+w+||\vv{B}||^2\right) \vv{U} -(\spr{U}{B})\vv{B}\right] =0 \,,\\
   \end{split}
\eeq
the Faraday equation
\beq
\pder{\vv{B}}{t}-\vv{\nabla}\!\times\!(\vpr{U}{B})=0\,,
\eeq
 and the differential constraint
\beq
\vdiv{B} =0\,.
\eeq
where $\rho$, $p$, $e_{th}(\rho,p)$ and $w(\rho,p)$ are the density, pressure, thermal energy and enthalpy of plasma respectively, $ \vv{U}$ is the fluid velocity and $\vv{B}$ is the magnetic field.    In this study, we use the equation of state for ideal gas with the ratio of specific heats $\kappa$, so $e_{th}=p/(\kappa-1)$ and $w=\kappa p/(\kappa-1)$, and use $\kappa=5/3$.

The simulations were carried out using a 3-rd order finite-difference scheme which is the Newtonian version of the scheme for ideal relativistic MHD described in \citep{KF25}, which in turn was inspired by the code ECHO \citep{DZ07}. For completeness, its key algorithms and some test simulations are given in Appendix \ref{sec:numerics}. 

In the setup of the simulations, all dependent variables are functions of only $x$ and $t$, so the wave vectors are aligned with the x axis, but the background (mean) magnetic field can be inclined to it.  In the equilibrium state, the mass density $\rho_0=1$, the x component of the magnetic field $B^x=B_0 =1$, and hence the Alfv\'en speed in the x direction is $c_0=B_0/\sqrt{\rho_0}=1$. For consistency with \citet{MT24}, in all simulations the magnetisation parameter  $b^2=a_0^2/c_0^2=0.1$, and hence $a_0=\sqrt{0.1}$ and $p_0=0.06$. The corresponding speeds of the slow and fast modes are  $c_{s,0}=0.22$ and $c_{f,0}=1.43$ respectively.  The wave vector of the mother wave points in the positive direction of the x axis, and to simplify both the simulations and the analysis of their results, the problems are set in the rest frame of the mother wave. Hence the x component of the flow velocity is $u_0^x=-1$.

\section{Monochromatic waves} 
\label{sec:mono-waves}

The case of monochromatic mother waves allows direct comparison of our simulations with the theory of parametric instability and previous numerical studies, and hence serves to increase the confidence in the results for wave packets.  Moreover, it lays foundation for their analysis. 

\subsection{Circularly-polarised wave}
\label{sec:cpmw}

\begin{figure}
\centering
 \includegraphics[width=0.8\columnwidth]{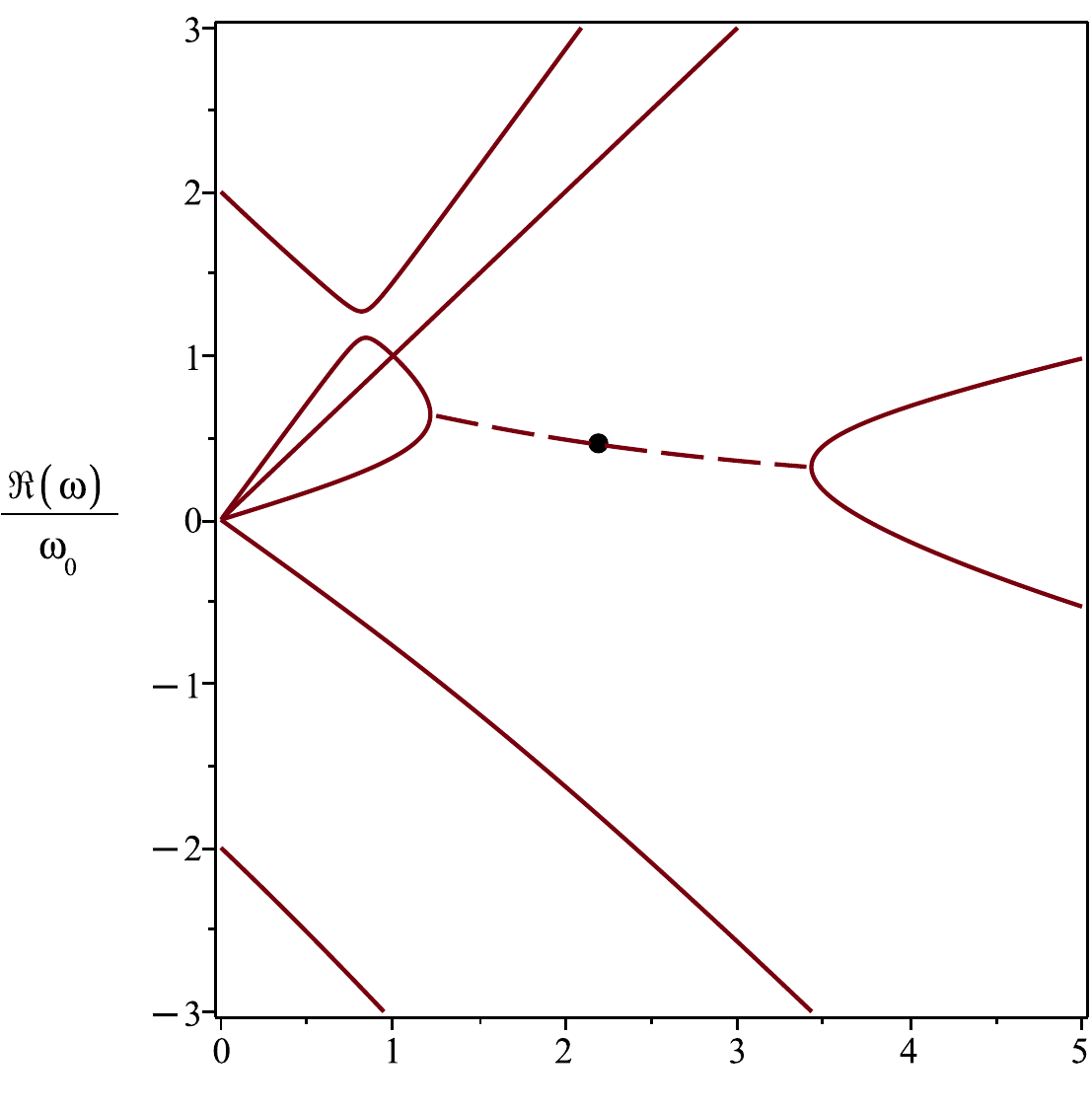}
 \includegraphics[width=0.8\columnwidth]{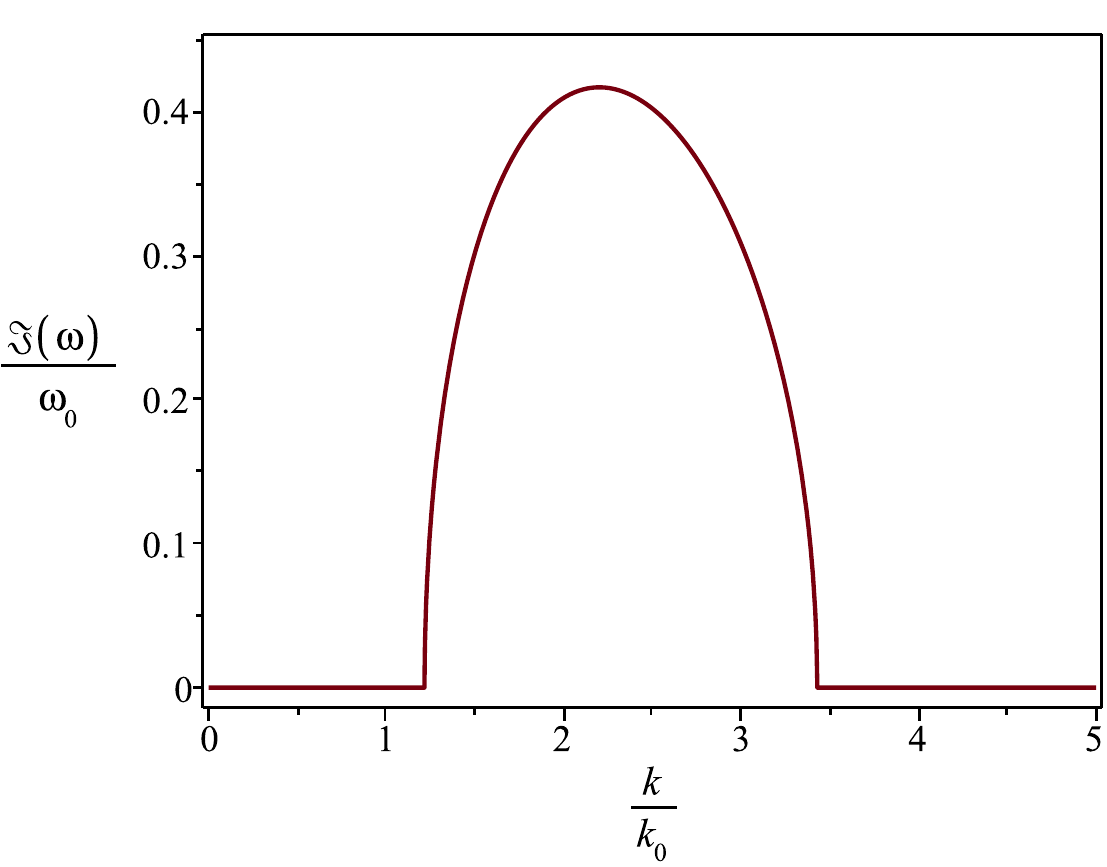}
\caption{Dispersion diagram for $b^2=0.1$ and $\eta=1$. Top panel: the real part of solutions to the dispersion equation \eqref{eq:dispersion} for $\tom$. The solid lines show the stable modes where the imaginary part of $\tom$ vanishes. The dashed line shows the real part of $\tom$ for two modes with complex conjugate values of $\tom$. The black dot on this line shows the location with maximal $|\gamma|$. Bottom panel: the positive imaginary part of the complex conjugate roots (the growth rate of the parametric instability) corresponding to the dashed line in the top panel.  }
\label{fig:disp}
\end{figure}

In the initial solution,  wave's magnetic field is 

\beq
\vv{B}_0=B_0\ort{x} + \eta B_0(\cos\phi(x)\ort{y} +\sin\phi(x)\ort{z})\,, 
\label{eq:cpmw-B}
\eeq
where $\phi(x)=k_0 x$,
and the fluid velocity 
\beq
 \vv{U}_0 = -\vv{B}_0/\sqrt{\rho_0} \,.
\label{eq:cpmw-u} 
\eeq
For consistency with \citet{MT24}, we use the same wave number $k_0=4$ and hence $\omega_0=1$. Figure \ref{fig:disp} shows the corresponding dispersion diagram. 
The computational domain is $[0,2\pi]$ with $n=942$ grid points and periodic BCs. 

Here, and in the rest of the problems, the equilibrium solution is perturbed via adding filtered flat white noise with the standard deviation $\sigma_\rho=10^{-5}$ to the plasma density.  To this aim, the FORTRAN subroutine RANDOM\_NUMBER($r$) is used to generate pseudo-random numbers $r_i$, $i=1\dots n$, one per each cell. Then each of them is converted into another random number $\bar{r}_i$ using centred mean filter with ($2m+1$)-point stencil.  Although not essential, this step allows to filter out high-frequency noise which otherwise causes an initial reduction in the perturbation amplitude due to numerical diffusion (See figure \ref{fig:dz-sg} for an example of such reduction.).  Finally, the initial perturbation is computed via

$$
\delta\rho_i = 2\sigma_\rho\sqrt{3(2m+1)} (\bar{r}_i-0.5) \,.
$$   
By trial and error, $m=8$ is found to be optimal. 

\begin{figure}
\centering
 \includegraphics[width=0.8\columnwidth]{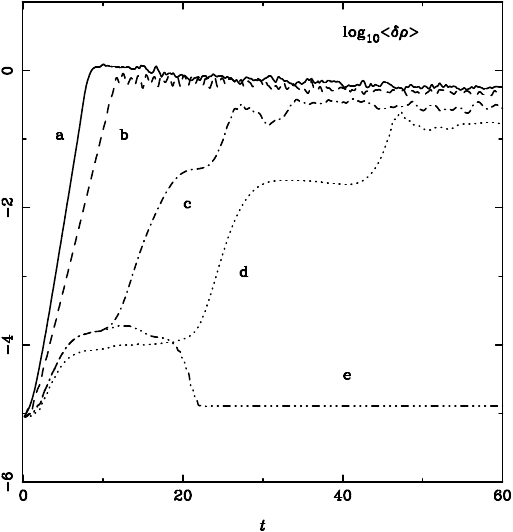}
\caption{The rms-value of the density perturbation as a function of time for a) the monochromatic circularly-polarised wave (solid line), b) the monochromatic arc-polarised wave (dashed line), c) the wave packet with $l/L=\pi/10$ and periodic BCs (dash-dotted line),  d) the wave packet with $l/L=\pi/20$ and periodic BCs (dash-dotted line), and e) the same as in c) but with open BCs (dash-tripple-dot line).  }
\label{fig:app-gam}
\end{figure}

The line `a' in figure \ref{fig:app-gam} shows the time evolution for the rms-value of $\delta\rho$ over the entire domain. At the linear phase of exponential grows, the two-point estimation of the growth rate yields $\gamma=1.64$ ($\gamma/\omega_0=0.41$).  

\begin{figure}
\centering
 \includegraphics[width=0.9\columnwidth]{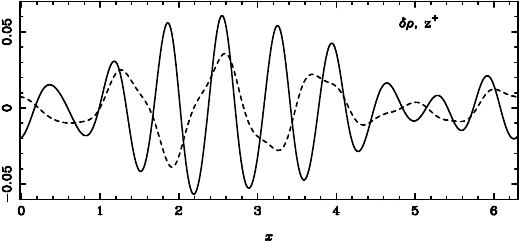}
 \includegraphics[width=0.9\columnwidth]{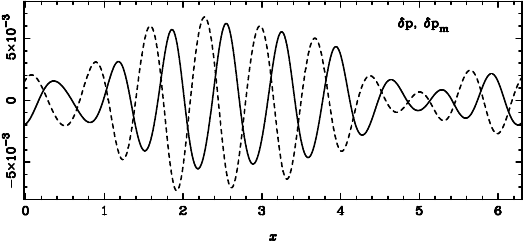}
\caption{Circularly-polarised monochromatic wave in the model with periodic BCs at $t=6$. Top pane: $\delta\rho$ (solid line) and $z^+$ (dashed line). Bottom panel: $\delta p$ (solid line) and $\delta p_m$ (dashed line)}
\label{fig:cpmw}
\end{figure}

Figure \ref{fig:cpmw} shows the numerical solution at $t=6$ for $\delta\rho$,  the z component of the Els\"asse variable $\vv{z}^+=\vv{B}+ \vv{U}/\!\sqrt{\rho}$, $\delta p$, and $\delta p_m$, where $p_m=||\vv{B}||^2/2$ is the magnetic pressure. The Els\"asse variable $\vv{z}^+$ is originally designed to detect normal reverse \A modes in uniform magnetic field. Although in this problem the transverse perturbations are not normal \A modes, the variable vanishes  in the unperturbed state, $\vv{z}^+_0=\vv{B_0}+ \vv{U_0}/\!\sqrt{\rho_0}=0$, and for this reason it is still well suited for detection of transverse perturbations.  

 Simply by counting the maxima (or minima)  of the curves, one finds that the spectrum of  the longitudinal perturbation peaks at $k_l=9$, and the spectrum of the transverse perturbations peaks at $k_t=5$. These satisfy the resonance condition $k_-=k-k_0$,  when $k$ is identified with $k_l$ and $k_-$ with $k_t$, suggesting that $z^+$ captures the sub-wave $\{\omega_-,k_-\}$. Plagging $\tk=9/4$ into the dispersion equation \eqref{eq:dispersion} yields the real frequency $\tom_r=0.457$ and the growth rate $\tilde{\gamma}=0.417$, which is in a good agreement with the value found in the simulations and is the same as the rounded to three significant digits maximum growth rate reached at $\tk_{max}=2.20$. The phase speed in the plasma frame $v_p=0.2c_0$ and so this is a forward wave. In the simulation frame, $v_p'=-0.8c_0$.
 
 Obviously, the longitudinal perturbations are not perfect harmonics, suggesting that several modes grow with similar rates.  The periodic BCs and the length of the computational domain restrict the problem to modes with integer wavenumbers. Of them, the closest two to $k=9$ ($\tk=2.25$) are $k=8$ ($\tk=2$) and $k=10$ ($\tk=2.5$). For the former, $\tilde{\gamma}=0.410$, and for the latter,  $\tilde{\gamma}=0.367$. 
 
 For $k_-=5$ ($\tk=5/4$), the corresponding frequency $\tom_-=\tom_r-1=-0.54$, and the phase speed in the plasma frame is $v_{p-}=-0.43c_0$, so this transverse sub-wave is a reverse one.  For the other transverse sub-wave, $\tk_+= 3.25$,  ($k=13$), $\tom_+=1.46$, and $v_{p+}=0.45c_0$. This sub-wave is a forward one. In the simulation frame, their phase speeds are $v_{p-}'=-1.43c_0$  and  $v_{p+}'=-0.55c_0$ respectively, and so both of them appear moving in the negative direction.   The ratio of real amplitudes of the transverse sub-waves $|\tilde{b_+}|/|\tilde{b_-}|=0.55$, and so the reverse sub-wave is only about twice as strong as the forward sub-wave.  This may explain the fact that the reverse wave dominates in the $z^+$ data, although $z^+$ is also naturally biased towards reverse waves.

\begin{figure}
\centering
 \includegraphics[width=0.9\columnwidth]{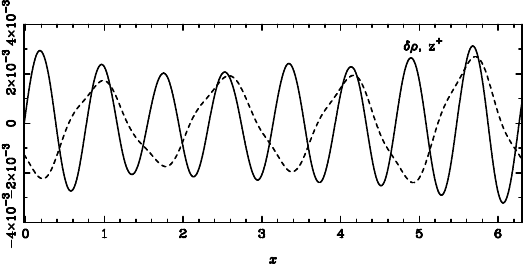}
 \includegraphics[width=0.9\columnwidth]{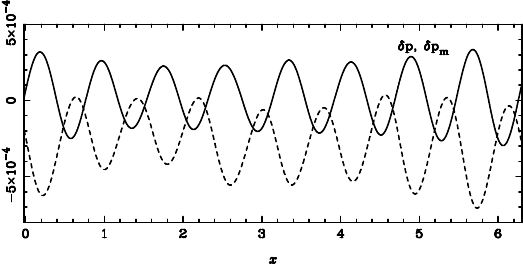}
\caption{Arc-polarised monochromatic wave in the model with periodic BCs at $t=6$. Top panel: $\delta\rho$ (solid line) and $z^+$ (dashed line). Bottom panel: $\delta p$ (solid line) and $\delta p_m$ (dashed line)}
\label{fig:apmw}
\end{figure}

\subsection{Arc-polarised wave}
\label{sec:apmw}

For the arc-polarised case, the magnetic field and velocity of the equilibrium state is still described by \eqref{eq:cpmw-B} and \eqref{eq:cpmw-u} , but now the phase function is   
\beq
\phi(x)=\sin (k_0 x) \,,
\eeq
with  $k_0=4$ as before.  This is the same phase function as used in \citet{MT24}. 

In figure \ref{fig:app-gam},  the evolution of the rms density perturbation in this model is shown by the curve `b'. At the linear phase its growth rate is $\gamma=1.07$ ($\gamma/\omega_0=0.27$). This value is in agreement with results obtained in \citet[][ see their figure 6 for $l/L \gg 1$]{MT24}.  
The same maxima counting as before yields $k_l=8$ and $k_t=4$ (see figure \ref{fig:apmw}). Although these wavenumbers differ from those found in the model with circular polarisation, they are still in agreement with the resonance condition $k_-=k-k_0$.  
By tracing the motion of individual peaks, the phase speeds in the plasma frame are $v_p=0.19$ for the longitudinal perturbation and $v_{p-}=-0.50$ for the transverse perturbation. These are very similar to the phase speeds found for the circularly-polarised case. Overall, the results show only slight changes compared to the case of circularly-polarised mother wave, which invites to look for at least an approximate analytic solution in this case too.

\section{Arc-polarised wave packets}
\label{sec:apwp}

\begin{figure}
\centering
 \includegraphics[width=0.8\columnwidth]{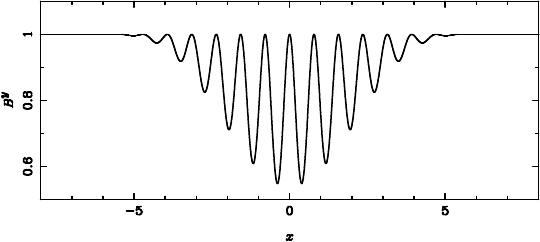}
 \includegraphics[width=0.8\columnwidth]{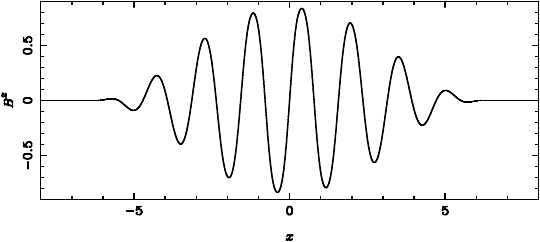}
  \includegraphics[width=0.8\columnwidth]{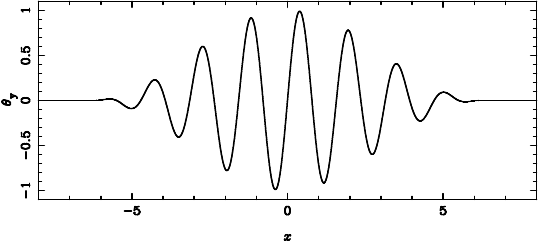}
\caption{Transverse magnetic field of the reference arc-polarised wave packet explored in the simulations. $\theta_y$ is the angle between this field and the y axis.}
\label{fig:awpack}
\end{figure}

The common way of introducing a wave packet via initial conditions involves multiplication of the monochromatic carrier wave $Ae^{ikx}$ by the amplitude-modulating ( envelope ) function $\chi(x)$, which reduces the wave amplitude to zero outside of the packet,  $f(x)=\chi(x)Ae^{ikx}$. However, Alfv\'en waves of compressible MHD keep invariant not only $u^x$, $\rho$ and $p$ but also the magnitudes of the magnetic field and velocity. So attempting to introduce an Alfv\'en wave-packet via $B_\perp=\eta B_0\chi(x) e^{ikx}$, like in \citet{LFD22}, would be a mistake.  
Instead, the modulation has to be applied to the wave phase \citep{MT24}.  The choice of modulating functions is wide. \citet{MT24} used the Gaussian envelope, $\chi(x)=\exp[-(x/l)^2]$, $x\in R$.  To prevent any potential complication in the case of  open BCs, we opted for the envelope with finite support 

\beq
  \chi(x)=\begin{cases} 
	\dfrac{(1+\cos k_ex)}{2} \,, &\text{if } -\pi/k_e \le x \le \pi/k_e \\
	0\,, &\text{otherwise } \\
\end{cases}\,.
\eeq 
Hence for the arc-polarised packet, the magnetic field is still given by \eqref{eq:cpmw-B}, but  the phase function is now

\beq
\phi(x)=\chi(x) \sin k_0x \,.
\eeq
For the main model, $k_e=k_0/8$, but models with smaller values are also explored in section \ref{sec:nbcs}.  This packet is illustrated in figure \ref{fig:awpack}. Some simulations, not presented in the paper, were carried out using the Gaussian envelope and their results are very similar.

\begin{figure*}
\centering
  \includegraphics[width=\columnwidth]{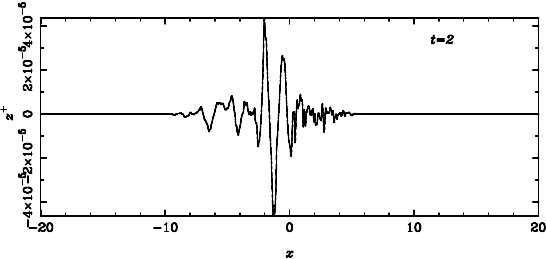}
   \includegraphics[width=\columnwidth]{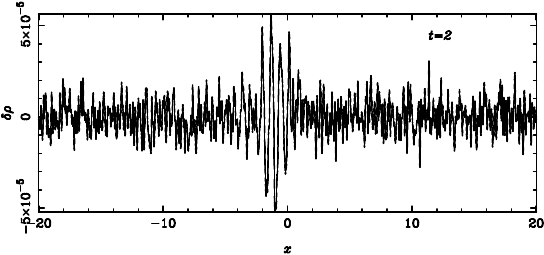}
  \includegraphics[width=\columnwidth]{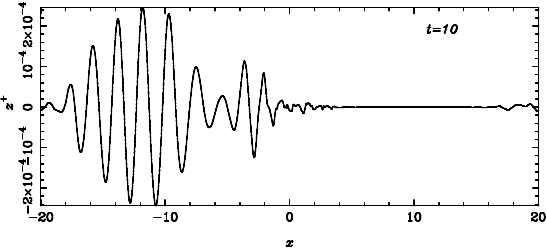}
  \includegraphics[width=\columnwidth]{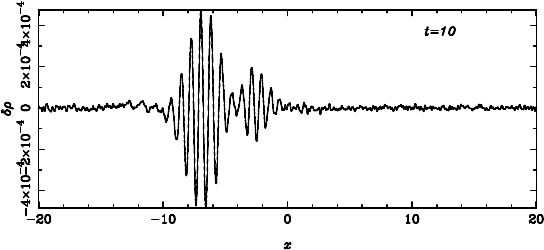}
  \includegraphics[width=\columnwidth]{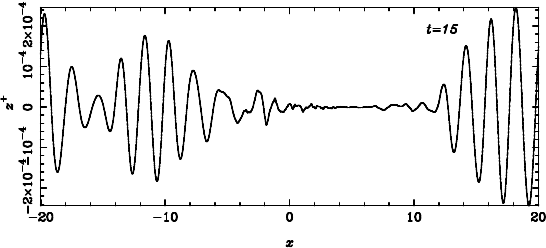}
  \includegraphics[width=\columnwidth]{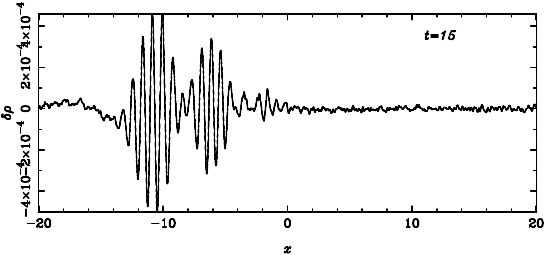}
  \includegraphics[width=\columnwidth]{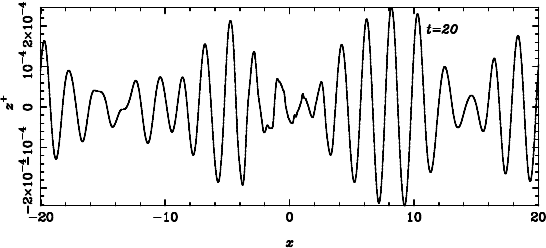}
   \includegraphics[width=\columnwidth]{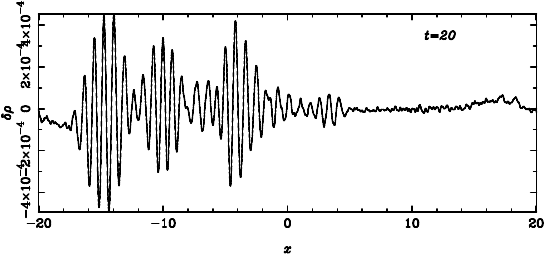}
  \includegraphics[width=\columnwidth]{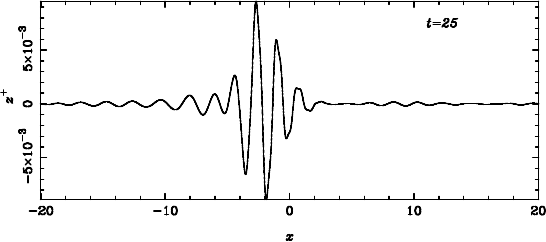}
  \includegraphics[width=\columnwidth]{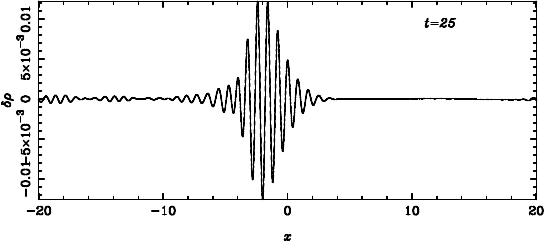}
\caption{Arc-polarised wave packet in the model with periodic BCs. Evolution of $z^+$ (left column) and $\delta\rho$ (right column).}
\label{fig:app}
\end{figure*}

\subsection{Models with periodic boundary conditions}

We started with the computational domain $[-10,10]$ with 3000 grid points and periodic BCs. The corresponding number of grid points per the wavelength of the carrier wave $\lambda_0=2\pi/k_0$ is almost the same as in the simulations of the monochromatic waves. The equilibrium is perturbed using the same algorithm as described in section \ref{sec:cpmw}. In figure \ref{fig:app-gam}, the growth of the density perturbation is shown by the curve `c'. The most interesting feature of the plot is that the growth rate experiences large variations, with intervals of relatively fast growth separated by intervals of much slower growth. Inspection of the numerical solution reveals that both the longitudinal and the transverse sub-waves do not come in the form of wave packets comoving with the mother-wave packet. Instead they are modulated waves travelling away from the mother packet in the downstream direction. They grow in amplitude while they are still inside the mother wave, but this growth terminates when they separate from it.  When they reemerge upstream due to the periodic BCs and meet the mother wave again, this triggers the next episode  of fast growth.    

To see this process clearly, we doubled the size of the computational domain, keeping the numerical resolution and the size of the mother packet the same. Since this increases the distance the daughter packets need to travel before meeting the mother wave for the next time by about 2.5 times, the expectation is that the duration of the low growth episodes increases by about the same factor.  This is confirmed by the simulations, as one can see in figure \ref{fig:app-gam}, where the evolution of the rms value for density perturbations in this model is shown by the line `d'.   
                   
\begin{figure}
\centering
 \includegraphics[width=0.9\columnwidth]{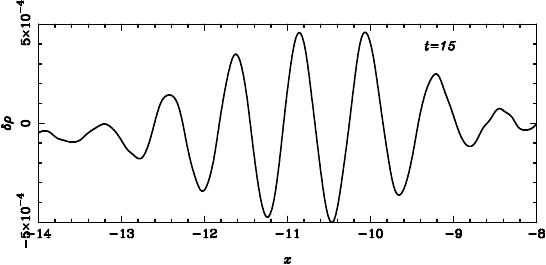}
  \includegraphics[width=0.9\columnwidth]{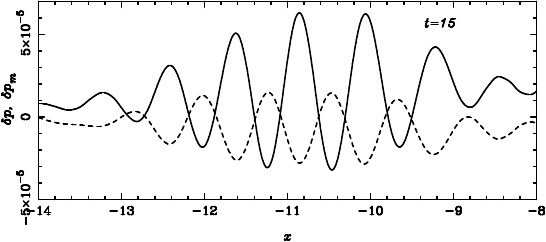}
\caption{Arc-polarised wave packet in the model with periodic BCs. Perturbations of the gas (solid line) and magnetic (dashed line) pressures accompanying  the density perturbations outside of the mother wave packet.}
\label{fig:nsw-pmp}
\end{figure}

\begin{figure}
\centering
 \includegraphics[width=0.9\columnwidth]{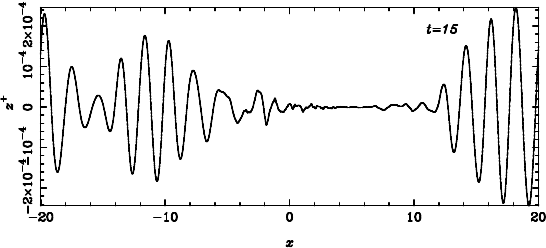}
  \includegraphics[width=0.9\columnwidth]{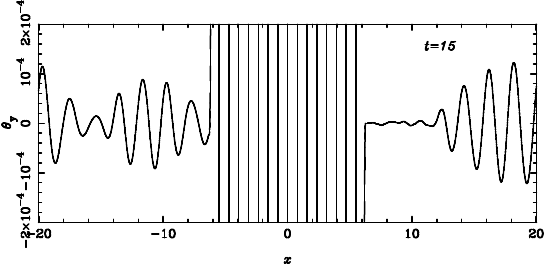}
\caption{Arc-polarised wave packet in the model with periodic BCs. Top panel: the z component of the Els\"asse variable  $\vv{z}^+$. Bottom panel: The angle $\theta_y$ between the transverse component of the magnetic field $\vv{B}_\perp$  and $\ort{y}$.}
\label{fig:naw-bt}
\end{figure}

Figure \ref{fig:app} illustrates the dynamics of the daughter waves from $t=2$ to $t=25$ covering the second half of the initial fast growth, the first episode of slow growth, and the first-half of the second episode of fast growth. Quick inspection of this figure shows that outside of the mother packet both the density and $z^+$ perturbations propagate preserving their shape, allowing to determine where each of the local peaks are later on with high degree of confidence. By measuring their displacement between $t=10$ and $15$, it is found that the phase speeds of the density and $z^+$ perturbations are $v_\rho'=-0.78$ and $v_{z^+}'=-2.0$ respectively. In the plasma frame, this corresponds to $\tilde{v}_\rho=c_s$ and $\tilde{v}_{z^+}=-c_0$. Thus these are the normal forward slow-magnetosonic and reverse Alfv\'en modes.  For the slow modes,  this is confirmed by the opposite variations of gas and magnetic pressures (see figure \ref{fig:nsw-pmp}).  For the Alfv\'en modes, by the lack of density variation (see figure \ref{fig:app}) and the variation of the magnetic field direction consistent with the variation of $\vv{z}^+$ (see figure \ref{fig:naw-bt}).

\subsection{Models with open boundary conditions}
\label{sec:openBCs}
The results of simulations with periodic BCs suggest that in simulations with boundary conditions allowing the daughter waves to escape the computational domain, the growth rate of the instability will be limited to the first episode of the fast growth seen in the simulation with periodic BCs. To check if this is the case, we repeated the simulations with the zero-gradient boundary conditions, where the values of all dependent variables in the first (and the last) domain cell are copied to the corresponding ghost cells.    These BSs, sometimes called `outflow', 'free-flow' or 'open' BCs, ensure high transparency of the domain boundaries to smooth waves.  

In figure \ref{fig:app-gam}, the growth of the density perturbation is shown by the curve `e'.  As one can see, for a while the growth of the rms value of $\delta\rho$ is indistinguishable from that in the model with the periodic BCs. However starting from $t\approx 10$, $\langle \delta\rho\rangle$ first stops growing and then drops down to approximately its initial value.      

\begin{figure}
\centering
  \includegraphics[width=\columnwidth]{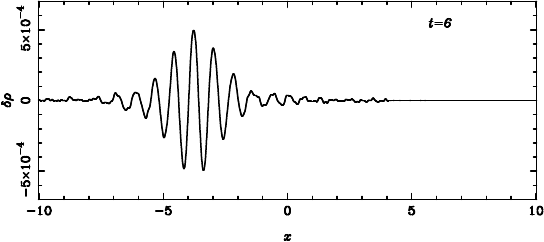}
   \includegraphics[width=\columnwidth]{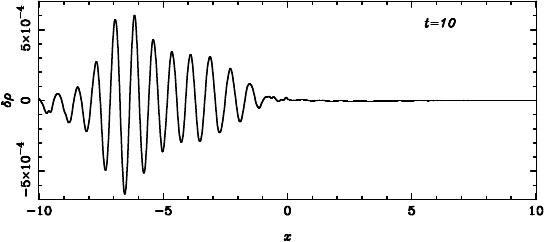}
  \includegraphics[width=\columnwidth]{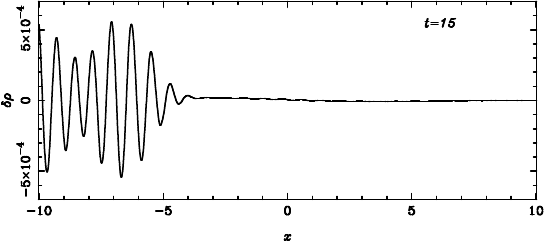}
  \includegraphics[width=\columnwidth]{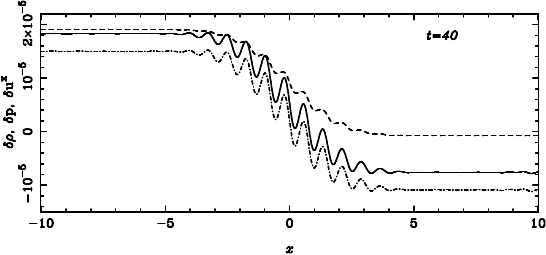}
\caption{Evolution of the density perturbation for arc-polarised wave packet with open BCs. At $t=40$ (the bottom panel) the solution has reached a steady-state. This panel also shows the gar pressure perturbation, $\delta p =p-p_0$ (dashed line), and the perturbation of the longitudinal component of velocity, $\delta u^x = u^x-u_0^x$ (dot-dashed line).   }
\label{fig:apf}
\end{figure}

Figure \ref{fig:apf} shows the evolution of the density perturbation in this model. By $t=6$ the small density perturbations initially occupying the whole domain are no longer present upstream of the mother packet.  They all have been advected by the incoming flow into the packet, leaving behind ($x>4$) and almost uniform flow.  Between $t=10$ and $t=15$ this lack of incoming perturbations leads to an overall decrease in the amplitude of the perturbations inside the mother packet ($-2\pi\!<\!x\!<\!2\pi$). By $t=20$, all the initially amplified perturbations have escaped from the domain via the downstream boundary, which results in the drop of $\langle\delta\rho\rangle$ seen in figure \ref{fig:app-gam}. What remains is the steady-state structure shown in the bottom panel of figure \ref{fig:apf}. It involves a monotonic density increase in the downstream direction combined with a spatial oscillation with the wave number $k=2k_0$.  Varying the numerical resolution shows that the amplitude of this structure behaves like $\propto\Delta x^3$, demonstrating that the structure is a numerical artefact created by the numerical diffusivity of our 3rd-order accurate numerical scheme.  

Thus the instability not just slows down but terminates.       

\subsection{Models with the noisy boundary conditions}
\label{sec:nbcs}

The results of the simulations with open BCs suggest that in the case of continuously incoming density perturbations, the mother packet will be followed by a trail of low-amplitude reverse Alfv\'en and forward slow waves. To generate the perturbations at the upstream (right) boundary, we use a stack of $m$ pseudo-random numbers $r_i$ obtained with the RANDOM\_NUMBER($r$) subroutine, and compute their mean value $\bar{r}$. It is used to set the random density perturbation with the standard deviation $\sigma_\rho$ via 
$$
\delta\rho_i = 2\sigma_\rho\sqrt{3m} (\bar{r}-0.5) \,.
$$    
At the next time step, the first random number of the stack is removed from it and a newly generated one is added to it, and a new value of the perturbation is computed in the same way.  This algorithm serves to filter out the high-frequency noise that is not amplified by mother wave but gets erased by the numerical diffusivity (In the simulations, $m=32$.). In addition, the x component of the ghost cells velocity is set $u^x=-c_0$, and  the zero-gradient BCs are applied to all other variables.  At the downstream boundary, the zero-gradient BCs are still applied to all variables. The computational domain is $x\in[-30,10]$ with 6000 grid points.

\begin{figure}
\centering
 \includegraphics[width=0.9\columnwidth]{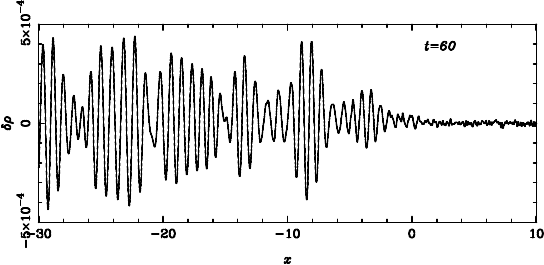}
  \includegraphics[width=0.9\columnwidth]{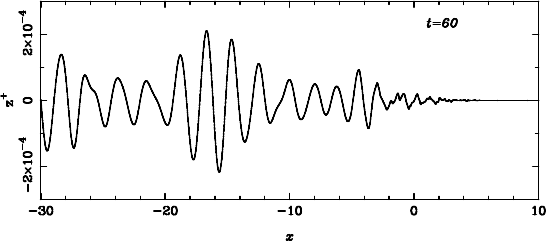}
\caption{Structure of the perturbations for the arc-polarised wave packet with in the model with noisy BCs at $t=60$ when the solution has reached a quasi-stationary state in the statistical sense.}
\label{fig:apnb-60}
\end{figure}

The results of simulation with such noisy BCs are consistent with the expectations - the mother wave does leave behind a trail in the form of modulated forward slow wave with the wave number $k_s\approx 7.5$ and modulated reverse Alfv\'en wave with the wave number $k_a\approx 3$ (see fig.\ref{fig:apnb-60}).   These wavenumbers approximately satisfy the resonance condition of the parametric instability $k_-=k-k_0$. 

For sufficiently long mother packets, one would expect the perturbations to become non-linear inside the packet. The critical length scale depends on the amplitude of incoming perturbations, the growth rate of the instability, and the phase speed of the unstable mode.  For long packets,  their inner structure is close to that of the unmodulated carrier wave, and one would expect  the instability growth rate to be similar to that of the monochromatic case. The same applies to the phase speeds of unstable modes. Since the modes of parametric instability are symbiotic, it is the speed of the fastest sub-wave, the reverse transverse sub-wave, that matters.  In the rest frame of the mother wave, its phase speed is of the order of the \A speed $c_0$ (see section \ref{sec:cpmw})\footnote{ \citet{Gold78} notes that this is likely to be the case in general, particularly for $\eta\ll 1$.}.   Taking the growth time as the \A crossing time of the packet $\tau_a=l/c_0$, where $l$ is the half length of the packet, the density perturbation reaches the density of the unperturbed state when

\beq
\sigma_{\!\rho} e^{\gamma\tau_a} =\rho_0\,,
\eeq  
where $\sigma_\rho$ is standard deviation of the incoming density noise,  and $\gamma$ is the instability growth rate for the carrier wave of the packet. Hence the critical length is
\beq
l_{max} = \frac{c_x\ln(\rho_0/\sigma_{\!\rho})}{\gamma} \,.
\label{eq:lmax}
\eeq 
Substituting the values  $c_0=\rho_0=1$ and $\sigma_{\!\rho}=10^{-5}$ used in our simulations, one obtains $l_{max}=11.5$. The actual half-length of the packet explored so far  is $l=\pi/k_e=2\pi <l_{max}$, consistent with the perturbations remaining linear, but only just.  

\begin{figure}
\centering
 \includegraphics[width=0.8\columnwidth]{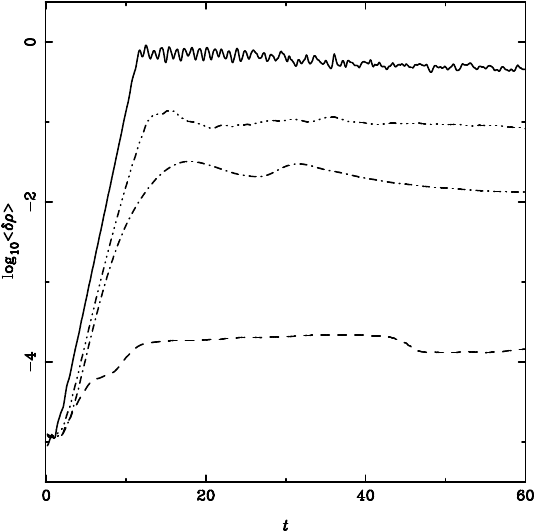}
\caption{Growth of rms $\langle\rho\rangle$ for wave packets of different lengths. The lines represent the solutions for arc-polarised carrier wave (solid line) and the wave-packets with $k_e=1/2$ (dashed line), $k_e=1/4$ (dash-dotted line) and $k_e=1/6$ (dash-triple-dotted line).} 
\label{fig:apnb-gam}
\end{figure}

To further verify \eqref{eq:lmax} and to explore the nonlinear regime we run additional simulations with $k_e=1/4$  ($l=4\pi$) and $1/6$ ($l=6\pi$).  Figure \ref{fig:apnb-gam} shows the growth of density perturbations in these models and in the case of monochromatic carrier wave.   It confirms that as $l$ increases the growth rate approaches that of the carrier wave, and so does the saturation level of perturbations.    
 
\begin{figure*}
\centering
 \includegraphics[width=\columnwidth]{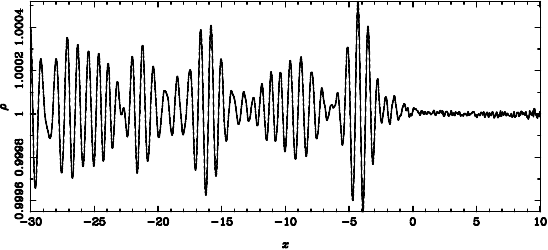}
 \includegraphics[width=\columnwidth]{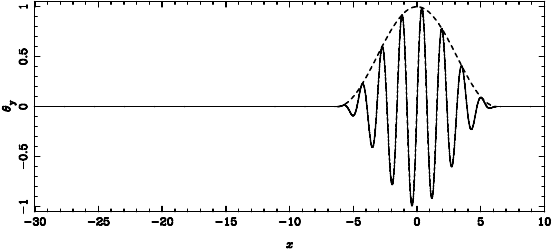}
 \includegraphics[width=\columnwidth]{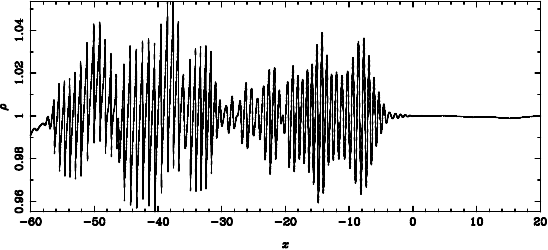}
 \includegraphics[width=\columnwidth]{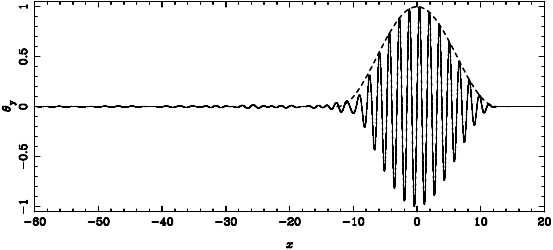}
 \includegraphics[width=\columnwidth]{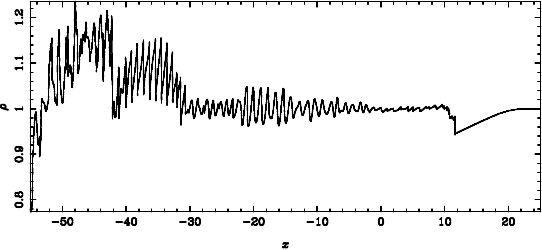}
 \includegraphics[width=\columnwidth]{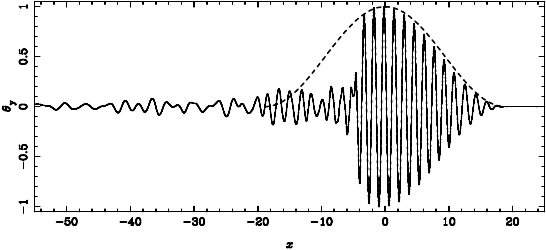}
\caption{The packet length effect on the solutions for arc-polarised wave packet in the model with noisy BCs at the phase of saturation ($t=70$). Left column: mass density $\rho$; Right column: the angle $\theta_y$ between the transverse component of the magnetic field and the y asix. The dashed lines in these plots show the initial envelopes of the mother packets. Top raw: $k_e=1/2$ ($l=2\pi$); Middlle raw:   $k_e=1/4$ ($l=4\pi$); Bottom raw:  $k_e=1/6$ ($l=6\pi$). Note the change in the domain size from $L=40$ in the top raw to $L=80$ in the other raws. }
\label{fig:apnb}
\end{figure*}

Figure \ref{fig:apnb} shows the numerical solutions for the three models at $t=70$,  well after the instability saturation point for all of them.  Whereas for $k_e=1/2$ and $k_e=1/4$ the instability saturates at linear amplitudes,   for $k_e=1/6$ it reaches the non-linear phase.  The most remarkable feature of the $k_e=1/6$ solution is the collapsed downstream section of the mother wave and its well-preserved upstream section. The half length of the preserved section is approximately 11, which is very close to the value of $l_{max}$ predicted by equation \eqref{eq:lmax}. The discontinuity in the density distribution seen at $x\approx 11$ is a forward fast shock, as evidenced by the correlated jumps in magnetic and gas pressures. It chases down a fast rarefaction occupying the section $x\in(11,20)$ and is chased by a train of weaker fast shocks occupying the section $x\in(0,11)$. The section with $x<-15$ is dominated by a train of forward slow shocks.    

\begin{figure}
\centering
 \includegraphics[width=\columnwidth]{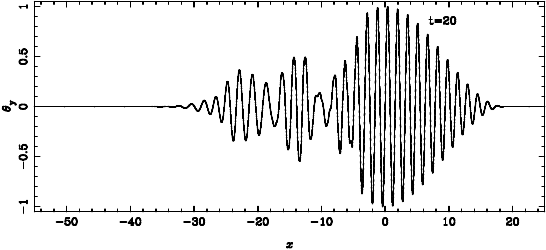}
  \includegraphics[width=\columnwidth]{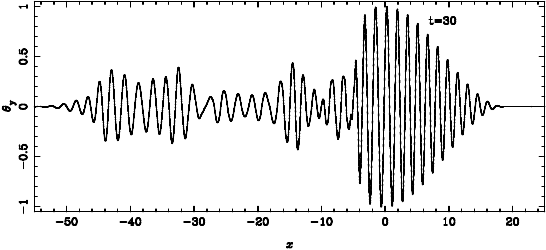}
   \includegraphics[width=\columnwidth]{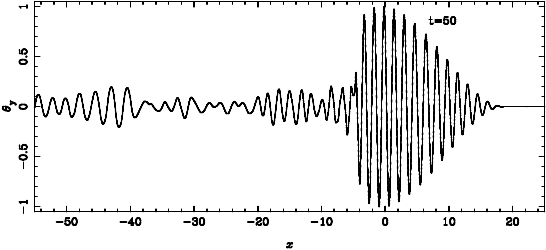}
    \includegraphics[width=\columnwidth]{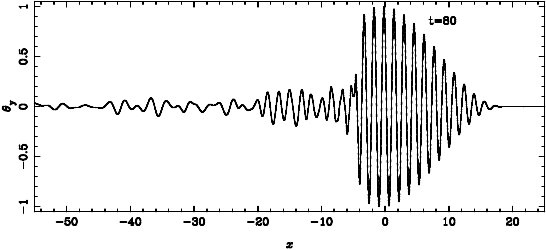}
\caption{Evolution of the mother packet in the model with $k_e=1/6$ . The angle $\theta_y$ at $t=80$ (solid line) and $t=50$ (dased line).}
\label{fig:apnb-ev}
\end{figure}

The dynamics of this collapse  is illustrated in figure \ref{fig:apnb-ev}. First a strong reverse \A wave is emitted by the mother packet, which is accompanied  by a reduction in the amplitude of packet's angular oscillations in its trailing section.  As the time increases, the amplitude of emitted reverse \A waves weakens and the packet enters the phase of slow evolutions.   It is hardly changed between $t=50$ and the end of the simulations at $t=80$.  
 
\section{Discussion}
\label{sec:discussion} 

Our results for arc-polarised Alfv\'en wave packets are in stark disagreement with the results of semi-analytical study by \citet{MT24}, which yields a temporal instability similar to the parametric instability of monochomatic Alfv\'en waves both for periodic and open BCs, though at a lower growth rate.   From the start of their analysis, they assume that in the frame of the packet, the perturbations have the form  $\delta f(x)e^{i\omega t}$, where both $\delta f(x)$ and $\omega$ are complex.  This describes a wave packet that neither changes its shape nor moves relative to the mother wave but only grows exponentially in amplitude. In contrast, the evolution of perturbations found in our simulations is much more dynamic and depends on the type of applied boundary conditions.  

With the periodic BCs, the perturbation grows intermittently.  Interacting with the initial density noise, the mother wave emits modulated forward slow magnetosonic and reverse Alfv\'en waves into its downstream. These waves then re-emerge upstream via periodic BCs and get amplified when they reunite with the mother wave.  Interestingly,  the mean growth rate decreases with the size $L$ of the computational domain, approximately like $\propto L^{-1}$, which is in agreement with the result by \citet{MT24}. However the reason for such behaviour is different. In our simulations, the mean growth rate  decreases with $L$ because the time required for the daughter waves to reunite with their mother grows like $\simeq L$.       

With the open BCs,  the initially generated daughter waves no longer reappear upstream but leave  the computation domain through the downstream boundary. The amplitude of perturbations inside the domain drops and the mother wave reaches a steady-state structure where $\rho$, $p$ and $u^x$ exhibit  spatial variations. These variations get reduces with increased numerical resolution at the same rate as the truncation errors of the scheme, implying that the structure is purely numerical and associated with the numerical viscosity and resistivity. When the density noise of the same standard deviation as the initial one is constantly generated at the upstream boundary, the solution does not settle to a steady-state, and the mother wave packet keeps emitting daughter waves. 

The reason why the daughter waves do not behave as comoving wave packets is connected to the reduction in the amplitude of the carrier wave in the wings of the mother packet. Whether it is modulated with the Gaussian function or our trigonometric function, at sufficiently large but finite distance from the packet centre its amplitude becomes so small that the carrier wave can no longer be considered non-linear.  When the total perturbation is small the general solution of MHD equations reduces to a mixture of non-interacting normal modes, and so the daughter waves become normal slow magnetosonic and Alfv\'en waves which freely escape from the mother packet given their negative phase speeds.

The mechanism of wave emission still appears to be related to the parametric instability, as the emitted waves satisfy the resonance condition $k_-=k-k_0$, though not exactly. Apparently, all the mother packets investigated in our study are sufficiently long for the eigenmode of the instability to get established and amplified in the central section of the packet, before splitting into normal MHD modes in its trailing wing.  

The results of all numerical experiments described in section \ref{sec:apwp} are consistent with picture where the perturbations enter the mother packet through its upstream wing and grow in amplitude as they move through the packet. If the packet is relatively short, the packet crossing time is also short and the perturbations remain linear when they escape from the packet through its downstream wing.  In this case, one would expect the packet to decay slowly, on the time scale greatly exceeding $1/\gamma$, where $\gamma$ is the growth rate of the parametric instability for monochromatic waves. Indeed, close inspection of the simulation data show no evidence of the packet decay between $t=20$ and $t=80$ for the models with $k_e=1/2$ and $k_e=1/4$ presented in section \ref{sec:nbcs}.   

When the length of the packet exceeds the critical value $l_{max}$ given by equation \eqref{eq:lmax}, the packet crossing time is long enough for the perturbations to reach non-linear amplitude. However, even in this case the upstream section of the packet with the length $l\lesssim l_{max}$ avoids destruction and presumably also enters the phase of slow decay. This may explain why arc-polarised \A wave packets are seen at large distances from the Sun \citep{Riley96}, where they are most likely generated.       
              
In many numerical studies of various fluid instabilities, there is no need to introduce initial perturbations or use noisy BCs because the noise associated with truncation or rounding errors does the job as well. So the fact that with open BCs the instability terminates and the mother packet settles to a steady state when the initial density perturbation stop coming in from the upstream (section \ref{sec:openBCs}) is somewhat curious.   As far as this limited effect of the initial noise is concerned, the result is in agreement with the understanding that the instability is not a stationary wave packet. It's growth is spatial and once the amplitude of the incoming noise drops, so does the amplitude of the perturbation inside the mother packet. For the same reason, the rounding errors play no role in the simulations - with the double precision computations its level is too low to be significant even if it is amplified while advected by the flow across the mother packet. The same likely applies to the noise that may emerge via the weak wave reflection off the open boundaries.         

It is somewhat less clear with respect to the truncation errors of the numerical scheme, which lead to the wavy steady-state structure in the distributions of density, pressure and longitudinal velocity inside the mother packet (the bottom panel of figure \ref{fig:apf}). Presumably, this wave does not have the properties required to trigger the instability. This wave is driven by the spacial variation of numerical dissipation associated with the oscillations of transverse magnetic field and velocity in the carrier wave (This is why its wave number $k=2k_0$.). This is not a compressive mode as it is `frozen' into the mother packet and hence moves with the \A speed $c_0$ through the plasma. The variations of $\rho$ and $u^x$ are also incompatible with the ideal compressive mode.            
          
 Our description of arc-polarised \A waves is different from that in a number of pioneering  works \citep{BH74,VH96,DZ01a}. 
 We simply set the radius of the arc and its angular size, whereas they first set the time-averaged (or phase-averaged) tangential magnetic field and then look for arcs that yield this value. With the additional assumption of sinusoidal variation for the component normal to the averaged field, the solution may not exist, and when it does, the angular size of the arc is below $\pi$. With our approach there are no such complications and one may choose an arc of any size. 
 
 A related issue is the type of polarisation an \A wave may have. In the publications related to the topics, in addition to circularly and arc-polarised waves,  there is mentioning of linearly-polarised \A waves, and spherically-polarised \A waves as well.    
 
 The term linear polarisation is an approximate description of \A waves with such a small angular size of their arcs that they are well approximated by a straight line. In this case, the variation of the transverse component of the magnetic field is small, and in this sense this is a linear wave. Nonlinear linearly-polarised \A wave is a contradiction in terms, as pointed out by \citet[e.g.][]{BH74}, but the notion of large-amplitude waves  involving variations of only magnitudes of transverse magnetic field and velocity persevere. This could be connected to the linearly-polarised waves of incompressible MHD originally discovered by Hannes \A \citep{{Alfven50}}, which propagate with the \A speed and have identical properties in both the linear and non-linear regimes. These solutions can be combined to yield a circularly- or arc-polarised mode,  and only this mode `survives' in compressible MHD as wave propagating with the \A speed, whereas the pure linearly-polarised modes turn into slow magnetosonic waves.  So in compressible MHD, initial solutions describing strong amplitude perturbations of transverse magnetic field and velocity inevitably move the problem under consideration in the category of arbitrary non-linear wave unidentifiable with any particular normal mode  \citep[e.g.][]{BH74,CH74}.   Slightly different is the degenerate case of vanishingly small transverse magnetic field. In this case, the \A speed becomes a double root of the dispersion equation (or equivalently a double eigenvalue in the eigenvalue problem for the  characteristic matrix of the compressible MHD equations), and the corresponding two-dimensional space of eigenvectors is the same as in incompressible MHD. Hence the small-amplitude waves are the same as in incompressible MHD and can describe both linearly- and circularly-polarised waves.              
 
The term ``spherical polarisation'' emerged from the observations of solar wind perturbations which keep the magnitude of the magnetic field fairly constant but change its direction in arbitrary way \citep{BD71,Balogh95}, with the arc-polarised events accounting to only up to 10\% of the data \citep{Riley96}.  Such behaviour  is inconsistent with the properties of individual \A waves in well defined background (mean) magnetic field and is indicative of strong \A turbulence instead \citep[e.g.][]{Barnes86}.  

Circular and arc polarisations are just special cases of the allowed variation of the transverse magnetic field and velocity in \A waves. In ideal MHD, the tip of the transverse magnetic field may travel along its circle in an arbitrary way, including jumps from one point to another. For example, it can make several full turns before jumping to another direction and tracing few small arcs.  

The origin of \A waves in the solar wind in general, and their arc-polarised instances in particular, is not well-established yet, although the dominance of modes moving away from the Sun in the plasma frame \citep{GSB95} indicates solar origin for most of them. The fluid motion on the Sun surface is dominated by turbulent convection and the associated motion of magnetic field foot-points must be quite random. So one would expect the \A waves generated there to be random from the very beginning \citep{CB05}.  This makes the simple models of parametric instability for monochromatic \A waves not particularly relevant.  In active regions, one may expect quasi-periodic oscillations of magnetic flux tubes associated with rapid restructuring of magnetic field via magnetic reconnection.  \citet{Jess_2009} reported possible quasi-periodic motion on the time-scale of hundreds of seconds and lasting few thousands of second above a group of bright spots associated with local concentration of magnetic field between convective granules of the Sun. Even if this type of phenomena may produce  \A wave packets, there is no physical reason for them to have a well defined frequency of the carrier wave.  \citet{Riley96} find a hint of the quasi-periodicity in one of the arc-polarised events in the solar wind, but even in this case the noise level is quite high.  So, future studies of the parametric instability should focus on random waves. Given the resonant nature of the instability, it is natural to expect a significant reduction of its efficacy for such waves.

\section{Summary}     

In this paper we described computer simulations of basic \A wave packets, constructed via modulation of monochromatic arc-polarised carrier wave, with the aim to establish the role of the parametric instability in their evolution. The initial setup utilises the inertial frame comoving with the packet, where the packet is an exact steady-state solution of ideal MHD. This steady state is perturbed via introducing small-amplitude density perturbations in the form of white noise passed through a low-pass filter to remove rapidly decaying via numerical diffusion high frequency waves.  Some of the simulations are carried out with the standard periodic and open boundary conditions, but for the most important cases the open boundary conditions at the upstream boundary are modified to turn them into a constant source of low-frequency noise.  In all these simulations, the plasma magnetisation is fixed to $a^2/c_0^2=0.1$, the transverse magnetic field strength parameter is fixed to $\eta=B_\perp/B_0=1$ and the amplitude of its angular oscillations is limited by $60^\circ$. 

All the results consistently point to developing of the parametric instability, but in the form of spatial growth of its daughter waves -  the incoming density perturbations trigger these waves in the upstream wing of the packet, which then grow in amplitude while propagating  through the mother wave towards its downstream wing and then escape the packet in the form of normal reverse \A and forward slow MHD waves. 

For sufficiently short packets, the perturbations emerge from the packet as small-amplitude waves. Their wave numbers are consistent with the resonance condition of the parametric instability.  The mother packet remains virtually unchanged on the e-folding time scale of the instability.       

For larger packets, the daughter waves reach non-linear amplitude while still inside the mother packet.  In this case, the growth rate of the instability at the linear phase is very close to that for the unmodulated carrier wave. At the non-linear phase,  the mother wave collapses starting from some position in the packet and then all the way to its downstream end.   However, its upstream section remains largely intact. The critical packet length separating the linear and non-linear regimes, as well as determining the size of the surviving section in the non-linear regime, can be estimated using the model of spacial growth based on the assumption that the e-folding length scale is given by the product of e-folding time scale for the unmodulated carrier wave and the \A speed.  

Ultimately, the simulations show that \A packets can survive for much longer than one would expect from the results for monochromatic \A waves, which may explain why such packets are found in the solar wind at large distances from the Sun.    

The results of our simulations are in conflict with the results of the semi-analytical study of arc-polarised Alfv\'en wave packets by \citet{MT24}, where the daughter waves are assumed to be wave packets comoving with the mother wave.

\section*{Data Availability}     

The data underlying this article will be shared on reasonable request to the corresponding author.

     

\bibliographystyle{mnras}
\bibliography{../../BibFiles/plasma,../../BibFiles/komissarov,../../BibFiles/numerics,../../BibFiles/AWaves,../../BibFiles/astro,../../BibFiles/physics}

\appendix

\section{Numerical scheme}
\label{sec:numerics}

Apart from the splitting algorithm, the numerical scheme used for this simulations is the same as in \citet{KF25} and borrows some key algorithms from the code ECHO developed by \citep{DZ07} for relativistic MHD equations. It is a 3rd-order accurate finite-difference scheme  employing the GLM approach,  3-rd order Runge-Kutter time integration, and a 3rd-order WENO reconstruction algorithm. 
 
The code integrates the evolution equations of MHD in the form conservation laws which can be combined into a single vector equation
\beq
\Pd{t}\vv{q}+\bmath{\nabla}\!\cdot\! \mathbf{F}=0 \,,  
\eeq
 where $\vv{q}$ is the vector combining the conserved variables, and $\mathbf{F}$ is the matrix combining corresponding flux tensors.
 
\subsection{GLM approach} 

To keep the magnetic field approximately divergence-free, we follow the method called Generalised Lagrange Multiplier \citep[GLM, ][]{Munz00,Dedner02}. Hence, we introduce an additional dependent scalar variable $\Phi$  and replace the Faraday equation and the divergence-free condition with 
\beq
 \Pd{t}\vv{B} + \vcurl{E}  +\vgrad\Phi =0 \,, 
 \label{eq:Faradays}
\eeq
\beq
   \Pd{t}\Phi +\vdiv\vv{B} =-\kappa\Phi  \,.
 \label{eq:divbs}
\eeq
In the simulations, $\kappa=0.2/\Delta t$, making the $e$-folding time for $\Phi $ (without the term $\vdiv\vv{B} $) equal to 5 integration time-steps $\Delta t$.  

\subsection{Time integration} 
\label{sec:TI}

Since this is a finite-difference scheme, the numerical solution $\vv{q}^n_{i,j,k}$ describes the values of $\vv{q}$ at the grid-points with coordinates $(x_i,y_j,z_k)$ at the discrete time $t^n$.  Here we utilise Cartesian coordinates and uniform spatial grid with $x_i=x_1 +(i-1) \Delta x$, $y_j=y_1 + (j-1) \Delta y $,  $z_k=z_1 + (k-1)\Delta z $, where $\Delta x=\Delta y=\Delta z=h$. These grid-points can considered as central points of rectangular computational cells with interfaces at $x_{i\pm1/2} =x_i \pm h/2$, $y_{j\pm1/2} =y_j \pm h/2$, and $z_{k\pm1/2} =z_k \pm h/2$. The time grid is also uniform, $t^n=t_0 +\Delta t n$ with $\Delta t = \mbox{C} h$, where C is the Courant number.  In the simulations $C=0.5$.

\begin{figure*}
\centering
 \includegraphics[width=0.3\textwidth]{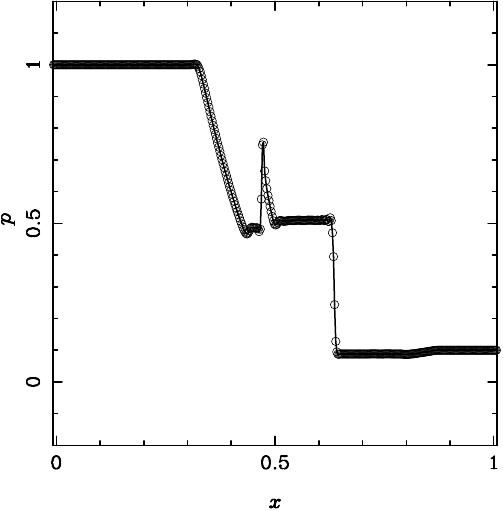}
  \includegraphics[width=0.3\textwidth]{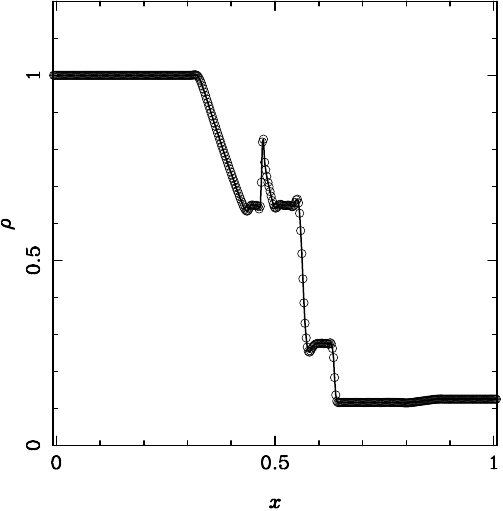}
  \includegraphics[width=0.3\textwidth]{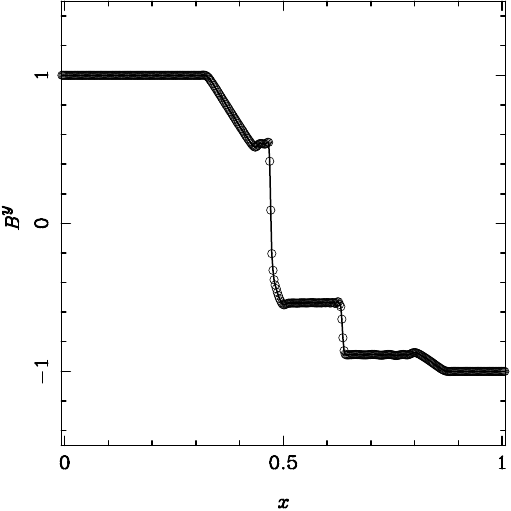}
\caption{Numerical solution to the \citet{BrWu88} test problem at $t=0.1$.}
\label{fig:brwu}
\end{figure*}

\begin{figure*}
\centering
 \includegraphics[width=0.3\textwidth]{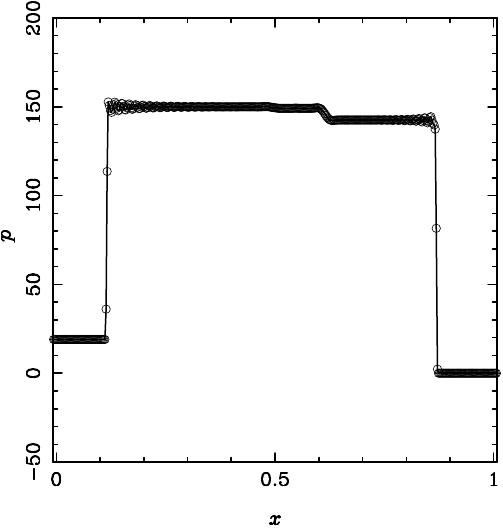}
  \includegraphics[width=0.3\textwidth]{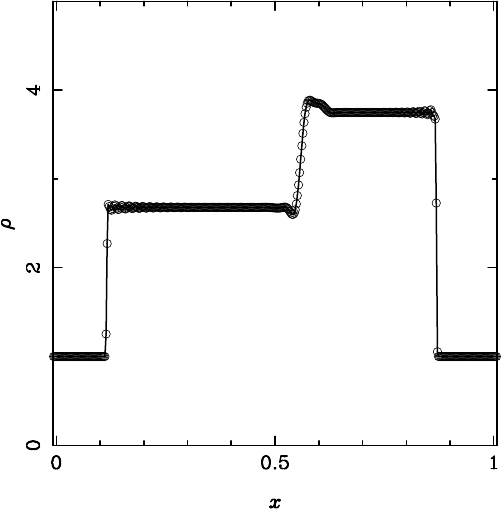}
  \includegraphics[width=0.3\textwidth]{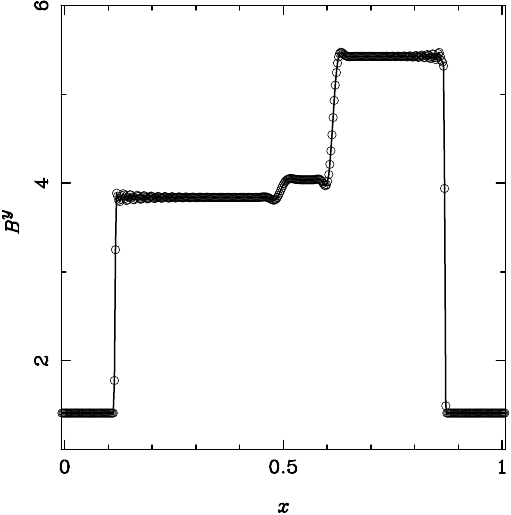}
\caption{Numerical solution to the \citet{RJ95} test problem at $t=0.08$.}
\label{fig:rj1}
\end{figure*}

The finite-difference equations have the form
\beq
\oder{\cal Q}{t}={\cal F}({\cal Q}) \,,
\label{eq:FDS}
\eeq
where ${\cal Q}$ is a one, two, or three dimensional  array, depending on the dimensionality of the problem. Each entry of this array is the vector $\vv{q}$  at the corresponding grid point. ${\cal F}$ is an array of the same dimension and size as ${\cal Q}$. Each entry of this array is the numerical finite-difference approximation for $-\vdiv{\vv{f}}+{\cal S}\sub{Q}$ at the corresponding grid point, where ${\cal S}\sub{Q}$ is the vector of source terms. In the case of Cartesian coordinates, the source terms emerge only in equation \eqref{eq:divbs}.   The system of ODEs \eqref{eq:FDS} is integrated using 3rd order Strong Stability Preserving (SSP) version of the Runge-Kutta method \citep{ShOs88}. Hence, 
\beq
{\cal Q}^{n+1} = {\cal Q}^n + \frac{\Delta t}{6} (k_1+k_2+4k_3) \,,
\eeq
where
\begin{align*}
k_1=&{\cal F}({\cal Q}^n)\,,\\
k_2=&{\cal F}({\cal Q}^n+\Delta t\, k_1) \,,\\
k_3=&{\cal F}\left({\cal Q}^n+\frac{\Delta t}{4} (k_1+k_2)\right) \,.
\end{align*}

The finite-difference approximation for $\vdiv{\vv{f}}$ is computed in the following steps: 
\begin{enumerate}
\item Conserved variables are converted into the primitive variables. This is needed because interpolating conserved variables may yield an unphysical  state.    

\item A 3rd order WENO interpolation \citep{KF25} is used to setup Riemann problems at the cell interfaces. 

\item HLL Riemann solver \citep{HLL} is used to find upwind flux densities $\vv{f}$ at the interfaces.  

\item  Central quartic polynomial interpolation is used to reconstruct the distribution of $\vv{f}$ in each coordinate direction and hence to find a 3rd-order approximation for  $\vdiv{\vv{f}}$  \citep[the DER operator in ][]{DZ07}.  This works fine for smooth solutions, but may introduce oscillations at shocks.  To avoid this, the computational domain is scanned for shock fronts and a 'safety zone' is set around them. Within the safety zone, a second-order TVD interpolation is used instead of the WENO interpolation.  
\end{enumerate}

The code was tested using a number of test problems. Here we present some of the 1D tests, including the parametric instability of monochromatic Alfv\'en waves with circular polarisation.   
      
\section{test simulations}

\subsubsection{Brio \& Wu problem}

This is a Riemann problem with the left state $p=1$, $\rho=1$, $\vv{u}=\vv{0}$, $\vv{B}=(0.75,1,0)$ and the right state  $p=0.1$, $\rho=0.125$, $\vv{u}=\vv{0}$, $\vv{B}=(0.75,-1,0)$.  The computational domain is $[0,1]$ with 400 grid points, and the initial discontinuity located at $x=0.5$. The numerical solution at $t=0.1$ is shown in figure \ref{fig:brwu}.  
This problem, which was first solved by \citet{BrWu88},  has become a must test for computational MHD in spite of the fact it involves a non-evolutionary intermediate shock which is super-Alfv\'enic relative to the upstream state and sub-Alfv\'enic relative to the downstream state. This property leads to its disintegration when it is approached from the downstream by an Alfv\'en wave \citep{fk-01}.     

\subsubsection{Ryu \& Jones problem}

This is a Riemann problem from \citet{RJ95} illustrated in their figure 1a.
Its left and right states are respectively 1) $p=20$, $\rho=1$, $\vv{u}=(10,0,0)$, $\vv{B}=(5B_0,5B_0,0)$  and 2)   $p=1$, $\rho=1$, $\vv{u}=(-10,0,0)$, $\vv{B}=(5B_0,5B_0,0)$, with $B_0=1/\sqrt{4\pi}$. The computational domain is $[0,1]$ with 400 grid points, and the initial discontinuity located at $x=0.5$. The numerical solution at $t=0.08$ is shown in figure \ref{fig:rj1}.

\begin{figure*}
\centering
  \includegraphics[width=0.3\textwidth]{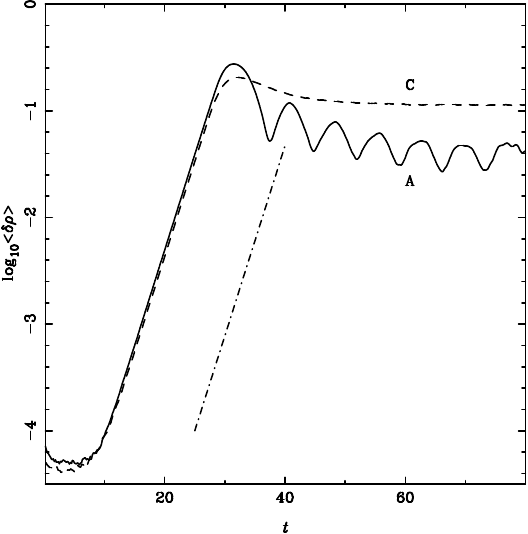}
 \includegraphics[width=0.3\textwidth]{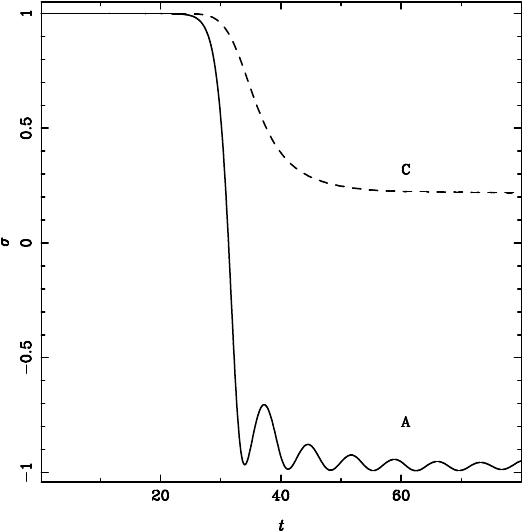}
\caption{Parametric instability of circularly polarized monochromatic Alfv\'en wave for the models A and C from \citet{DZ01}. Left panel: Growth of $\langle\rho\rangle$. The dash-dotted line shows exponential growth with the rate $\gamma=0.41$ ($\gamma/\omega_0=0.10$). Right panel:  Evolution of parameter $\sigma_{\cE}$. }
\label{fig:dz-sg}
\end{figure*}
\begin{figure}
\centering
 \includegraphics[width=\columnwidth]{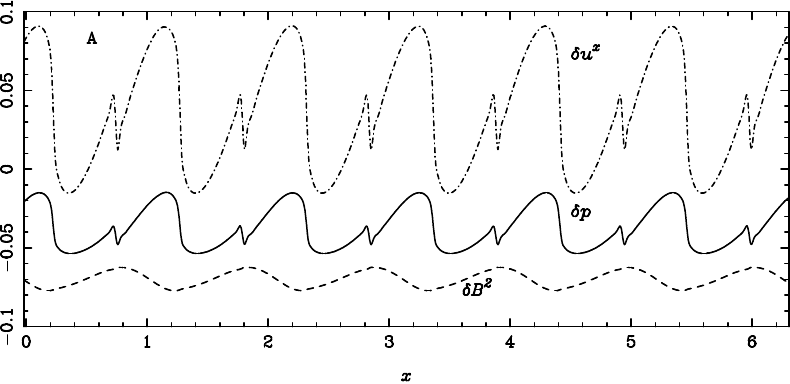}
 \includegraphics[width=\columnwidth]{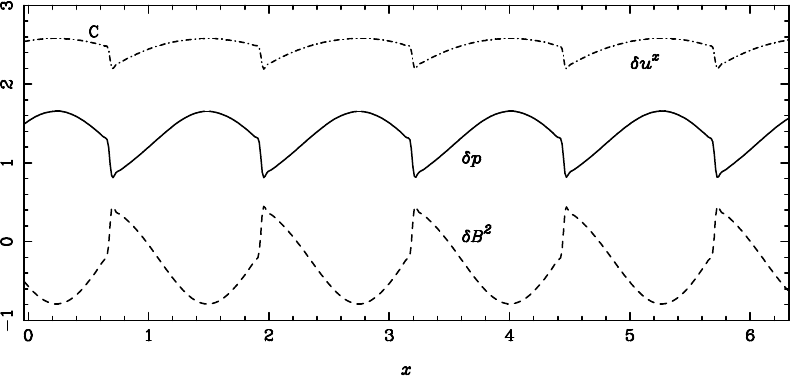}
\caption{Parametric instability of circularly polarized monochromatic Alfv\'en wave.  Top panel: Model A from \citet{DZ01}. Numerical solution at $t=40$. Bottom panel: Model C from \citet{DZ01}. Numerical solution at $t=40$}
\label{fig:dz-40}
\end{figure}

\subsubsection{Parametric instability of monochromatic circularly-polarised Alfv\'en waves}

Since the topic of this study is the stability of large-amplitude Alfv\'en waves, we repeated the 1D simulations of \citet{DZ01}  of monochromatic circularly-polarised Alfv\'en waves in a domain with periodic boundary conditions, namely their models A and C.  We used both the original setup of \citet{DZ01}, where $u_x=0$, and the setup in the rest frame of the mother wave. For both the setups, the results are almost identical to the original.   

Here we present the results for the setup in the rest frame of the mother wave.  In the initial solution,  $\rho_0=1$,  
\beq
\vv{B}_0=\eta B_0(1,\cos(k_0x),\sin(k_0x))\,,\quad    \vv{U}_0=-\vv{B}_0/\sqrt{\rho_0}\,,
\eeq
with  $B_0=1$, $k_0=4$, and $p_0=({\beta_0}/{\gamma}) B_0^2$ with $\gamma=5/3$ for both these models. In the model A, $\beta=0.1$ and $\eta=0.2$, whereas in the model C,  $\beta_0=1.2$ and $\eta=1$. The computational domain is 
$[0,2\pi]$ with 942 computational cells and periodic boundary conditions. This equilibrium solution is perturbed via adding unfiltered flat white noise to $\rho_0$ with the mean deviation $\sigma_\rho=10^{-4}$.

Figure \ref{fig:dz-sg} shows the time evolution of the rms value of density perturbation $\langle\delta\rho\rangle$ and the cross-helicity 

\beq
\sigma_{\cE} =\frac{\cE^+ -\cE^-}{\cE^+ +\cE^-} \,, 
\eeq  
where $\cE^\pm=\langle||\vv{z}^\pm||^2 \rangle$ are the Els\"asser energies of the forward and reverse Alfv\'en waves and $\vv{z}^\pm=\delta\vv{u} \pm \delta\vv{B}/\!\sqrt{\rho}$. These are in agreement with the results by \citet{DZ01} presented in their figure 2 down to fine details. The same is true for the solutions at the non-linear stage ($t=40$) of the instability shown in figure \ref{fig:dz-40} of this paper and in the figure 3 of \citet{DZ01}.  

\end{document}